\documentclass[aps,preprint]{revtex4}%
\usepackage{amsfonts}
\usepackage{amsmath}
\usepackage{amssymb}
\usepackage{graphicx}%
\begin{document}
\preprint{ }
\title[Novel numerical method]{A Novel Method for the Solution of the Schr\"{o}dinger Eq. in the Presence of
Exchange Terms.\bigskip}
\author{George H. Rawitscher,}
\affiliation{Physics Dept., Univ. of Connecticut, Storrs CT 06268}
\author{S. -Y. Kang and I. Koltracht,}
\affiliation{Mathematics Dept., Univ. of Connecticut, Storrs CT 06268}
\author{Essaid Zerrad and Kamal Zerrad,}
\address{Dept. of Physics and Pre-engineering, \linebreak\\
Delaware State University, Dover, Delaware,}
\author{B. T. Kim,}
\address{Dept. of Physics and Institute for Natural Sciences,\\
Sungkyunkwan Univ., Suwon 440-74,}
\author{T. Udagawa,}
\affiliation{Dept. of Physics, Univ. of Texas, Austin, TX 78712}
\author{}
\pacs{PACS number}
\pacs{PACS number}
\pacs{PACS number}

\begin{abstract}
In the Hartree-Fock approximation the Pauli exclusion principle leads to a
Schr\"{o}dinger Eq. of an integro-differential form. We describe a new
spectral noniterative method (S-IEM), previously developed for solving the
Lippman-Schwinger integral equation with local potentials, which has now been
extended so as to include the exchange nonlocality. We apply it to the
restricted case of electron-Hydrogen scattering in which the bound electron
remains in the ground state and the incident electron has zero angular
momentum, and we compare the acuracy and economy of the new method to three
other methods. One is a non-iterative solution (NIEM) of the integral equation
as described by Sams and Kouri in 1969. Another is an iterative method
introduced by Kim and Udagawa in 1990 for nuclear physics applications, which
makes an expansion of the solution into an especially favorable basis obtained
by a method of moments. The third one is based on the Singular Value
Decomposition of the exchange term followed by iterations over the remainder.
The S-IEM method turns out to be more accurate by many orders of magnitude
than any of the other three methods described above for the same number of
mesh points.

\end{abstract}
\date[Date text]{date}
\date{}
\date[Date text]{date}
\date{}
\startpage{1}
\endpage{ }
\maketitle
\tableofcontents

\section{\bigskip Introduction.}

A new very efficient and stable method for solving the Schr\"{o}dinger
equation has recently been developed \cite{IEM}. This method (S-IEM) solves
the Lippman-Schwinger integral equation rather than the differential
Schr\"{o}dinger equation because the numerical errors for the solution of the
former are inherently smaller than for the latter, and by making a spectral
expansion of the solution into Chebyshev polynomials it acquires additional
excellent accuracy properties. This method also avoids the usual drawback of
numerical solutions of integral equations, namely, the need to invert large
non-sparse matrices which represent the discretized form of the equation. It
achieves this by dividing the integration interval into partitions and by
making use of the semi-separable nature of the Green's function in
configuration space. The accuracy of the S-IEM has been tested by comparing it
to the solution of the differential Schr\"{o}dinger equation for various
cases, such as the scattering of cold atoms \cite{IEMA}, and for tunneling
through a barrier \cite{MORSE}, but not by comparing it to existing solutions
of the integral Lippman-Schwinger equation. The purpose of the present paper
is two-fold: a) to extend the S-IEM to the case that there are non-local
exchange terms present. This is possible \cite{baps}, \cite{KG} without losing
the original accuracy and stability features since the kernel of the exchange
integral is semi-separable (see below). And b), to compare the accuracy of the
extended S-IEM to conventional methods of solving an integral equation in
configuration space.

The scattering of an electron from a hydrogen atom offers a good opportunity
for performing such a test, because, integral equation methods have been
developed in the past in order to include the integral exchange terms required
by the Pauli exclusion principle, as is described further below. The present
test calculation is physically not realistic since it does not allow for the
polarization of the bound electron cloud by the incident electron, or the
ionization of the atom. But it is still sufficiently close to realistic so as
to serve as an adequate test for the comparison of different algorithms.
Realistic calculations which allow for the polarization of the electron cloud
involve coupling between many channels \cite{BRAY}, or else the solution of a
two-dimensional differential equation \cite{STELBO}, which is much beyond the
scope of the present work, and is not necessary in order to demonstrate the
power of our new method.

The three methods for solving the non-local Schr\"{o}dinger equation in the
presence of the exchange terms we compare our S-IEM with\ are as follows. One
of these methods \cite{kouri} solves an integral equation which is very
similar to the one we solve, with the main difference that it uses a trapezium
type method for numerically expressing the integrals, rather than the spectral
Chebyshev method used in the S-IEM. Another is an iterative method introduced
by Kim and Udagawa in 1990 \cite{udagawa}\ for nuclear physics applications,
which makes an expansion of the solution into an especially favorable basis
obtained by a method of moments. The third is a variation of a conventional
iterative method for including the exchange term. It consists in expanding the
exchange kernel into a small number of separable terms by means of the
Singular Value Decomposition and then iterating over the remainder
\cite{essaid}. Since there is a rich literature on methods developed for
taking the exchange terms into account, a review of some of the methods most
relevant to the S-IEM will be described below.

The scattering of electrons from atoms or molecules has been the subject of
investigation ever since quantum mechanics was introduced, and the research
continues unabated, mainly concerning the scattering of polarized electrons
\cite{KLEINP} or of high energy photons \cite{PRATT} from atoms or molecules,
or charge transfer in ion-molecule scattering \cite{SAHA}. The latter can also
be calculated by using advanced three-body formalisms \cite{ALT}. One of the
features which gives computational difficulties in solving the corresponding
Schr\"{o}dinger equation is the Pauli exclusion principle which requires that
the wave function of the incident electron be anti-symmetric with the wave
functions of the electrons in the target atom or molecule. In the
Hartreee-Fock formulation this requirement leads to the presence of non local
terms in the (coupled) differential equations of the form
\begin{equation}
\int_{0}^{\infty}K(r,r^{\prime})\,u(r^{\prime})\,dr^{\prime}, \label{nl}%
\end{equation}
where $u(r)$ is the required solution and $K$ is the integration kernel due to
exchange. In the early investigations \cite{burke1} the exchange terms were
taken into account iteratively, by using Green's functions defined by the
local part of the potential. However, under certain conditions these
iterations do not converge \cite{burke1}, \cite{TEMKIN}. Methods to accelerate
the convergence have been introduced, but such methods tend to be cumbersome
and unpredictable. An improved iteration procedure can be obtained by means of
a separable representation of the integral kernel\cite{essaid}, as is
described further below, but this method has its limitations as well.
An\ interesting method to include the exchange terms non-iteratively by
approximating them in terms of a separable representations has been developed
by Schneider and Collins \cite{SCsep}. Methods to solve the non-local
Schr\"{o}dinger equation non iteratively and rigorously have also been
developed \cite{NIEM}. One of the oldest ones originated with I. Percival and
R. Marriott \cite{MARRIOT}. It consists in constructing auxiliary functions
which are solutions of the differential equation in the presence of several
types of inhomogeneous terms and then constructing the exact solution (which
take into account the integral exchange terms) by means of linear combinations
of the auxiliary functions. A variation of this method was developed by Lamkin
and Temkin \cite{TEMKIN} for their Polarized Orbitals procedure. In this
method the semi-separable nature of the exchange kernel $K$
\begin{align}
K(r,r^{\prime})  &  =A(r)\,B(r^{\prime})\;\;for\;\;r^{\prime}<r\nonumber\\
K(r,r^{\prime})  &  =A(r^{\prime})\,B(r)\;\;for\;\;r^{\prime}>r \label{Kss}%
\end{align}
is exploited. The integral of Eq. (\ref{nl}) then becomes
\begin{equation}
\int_{0}^{\infty}K(r,r^{\prime})\,u(r^{\prime})\,dr^{\prime}=A(r)\,\int
_{0}^{r}B(r^{\prime})\,u(r^{\prime})\,dr^{\prime}+B(r)\,\int_{r}^{\infty
}A(r^{\prime})\,u(r^{\prime})\,dr^{\prime} \label{Is}%
\end{equation}
which can be rewritten as
\begin{equation}
\int_{0}^{\infty}K(r,r^{\prime})\,u(r^{\prime})\,dr^{\prime}=A(r)\,\int
_{0}^{r}B(r^{\prime})\,u(r^{\prime})\,dr^{\prime}-B(r)\,\int_{0}%
^{r}A(r^{\prime})\,u(r^{\prime})\,dr^{\prime}+CB(r). \label{IV}%
\end{equation}
Again auxiliary functions can be obtained by solving the differential integral
equation with the constant $C$ set equal to zero or unity. The true solution
$u(r)$ is obtained by means of a linear combination of the auxiliary
functions, and the constant $C=\int_{0}^{\infty}A(r^{\prime})\,u(r^{\prime
})\,dr^{\prime}$ as well as the coefficients of the linear combination can be
obtained via the solution of an algebraic equation. The advantage of this
method \cite{TEMKIN} is that, when the integral over the kernel goes from $0$
to $r,$ as is the case in Eq. (\ref{IV}), the solution of the
integro-differential equation can be performed easily by starting at the
origin and increasing the upper limit gradually from one meshpoint to the
next. With a step size of 0.05 the authors obtain accuracies to about three
significant figures with this method.

A method which yields five significant figures with the same step size of 0.05
is obtained by Kouri and co-workers \cite{kouri}. The main difference from
Temkin's method \cite{TEMKIN} is that the authors first transform the
integro-differential equation into a Lippman-Schwinger integral equation,
because integral equations have greater numerical stability than differential
equations. They again transform the integrals, which originally extend from
$0$ to $\infty$ into integrals from $0$ to $r,$ plus inhomogeneous terms, thus
obtaining Volterra integral equations of the second kind. They then easily
obtain auxiliary solutions to auxiliary Volterra equations by stepping
progressively from the origin to increasing values of $r,$ similarly to what
is done in the method of Temkin .The exact solution, and the respective
constants, can then be determined in a algebraic way similar to what is done
in Ref \cite{TEMKIN}. Smith and Henry \cite{NIEM} have also developed methods
to solve the Volterra type integral equations non iteratively. These methods
are generally called NIEM, where the ``N'' stands for ``Non-iterative''.
Collins and Schneider also have examined the NIEM form of the non-local
Schr\"{o}dinger equation, \cite{CSint} but without transforming it into a
Volterra type. By discretizing the integral via the trapezoidal rule, they
obtain a linear algebraic equation for the wave functions at the mesh points,
and for this reason the method is called (LA). The formulation of the initial
equation to be solved by our new method is very similar to that of
\cite{CSint}. The main differences, to be discussed below, arise from the
numerical techniques used in the solution of these equations (NIEM). For the
case of the exchange terms which are due to the Coulomb interaction, as is the
case for most atomic physics calsulations, the exchange terms can also be
replaced by coupling to a set of ''pseudo-states, a technique which is made
use of in the work of Weatherford, Onda and Temkin \cite{WOT}, and is also
illustrated in the present work.

Other methods have been developed to solve the electron-atom scattering
equations. One consist in introducing a set of basis functions such as
Laguerre polynomials, and expanding both the solution and the target states
into this basis \cite{exp}. Another such expansion basis utilizes sturmian
functions\cite{sturm}. In these procedures the exchange integrals can be
carried out, since they contain the known basis wave functions. Finite element
representations have also been applied \cite{shertz}. The R-matrix approach is
also well developed \cite{R-M}\ .

As already briefly mentioned above, our new ``spectral integral equation
method'', S-IEM, transforms the differential equation into an equivalent
Lippmann-Schwinger integral equation through the use of Green's functions,
similar to what is done in the older approaches described above. It differs
from the older methods in that it uses the Fredholm form of the integral
equation (whose range of integration is from $0$ to $r_{\max}$ ), as done in
Ref. \cite{CSint}, and does not transform it into a Volterra type (whose range
of integration is from $0$ to the variable radial distance $r$). It thus it
avoids the need to evaluate the constants $C$ which occur in theVolterra
method, but instead it has to solve for a larger number of other constants.
The latter arise by dividing the radial integration interval $[0$ $r_{\max}]$
into partitions, and by expanding the solution\ in each partition into two
independent functions which in turn are obtained by solving a local integral
equation through the spectral expansion into Chebyshev Polynomials. There are
twice as many such coefficients as there are partitions, and hence the matrix
from which the coefficients are calculated is large, say $600\times600$.
However, this matrix is sparse, and hence soluble economically. As is shown
here, the semi-separable structure of the exchange nonlocality allows us to
preserve the sparseness in the present case as well. If, however, the nonlocal
potential is not of the semi-separable form our integral equation method still
gives stable and accurate solutions, but then the biggest matrix involved is
no longer sparse \cite{KKR}. From the numerical point of view, the main
difference of our S-IEM from other integral equation methods described above
is that the radial mesh-points in the S-IEM are not equidistant, while those
for the latter are. The numerical errors of the latter are of the finite
difference type, and are given by a fixed power of the distance between mesh
points, while in the S-IEM the errors become smaller than any power of the
distance between mesh points. Or, more precisely, in the S-IEM the errors
become smaller than any inverse power of the number of Chebyshev support
points in each partition, a property which expresses the spectral nature of
the S-IEM. This property permits the S-IEM to have far fewer mesh points than
the more conventional discretization formulations of differential or integral
equations for a given envisaged accuracy, as will be demonstrated in Fig.
\ref{FIG4} below.

The first of our three comparison methods \cite{essaid} consists in replacing
the exchange kernel by a small number of fully separable terms, and carrying
out iterations only over the remainder. This is possible because, as is well
known, the Green's function for a Schr\"{o}dinger equation with both local and
non local but fully separable potentials can be obtained without much
difficulty by adding terms to the Green's function distorted only by the local
potential. Our second method \cite{udagawa} is a modified integral equation
method, denoted as M-IEM. It uses a set of basis functions which are obtained
by applying successively higher powers of the hamiltonian operator with local
potentials on an initial scattering wave function. This method has been very
successful in applications to nuclear physics problems. The third method
\cite{IEM}, the S-IEM, uses the Lippmann-Schwinger integral form of the
Schr\"{o}dinger equation. It differs from a previously introduced
non-iterative solution of the Lippmann-Schwinger integral equation, denoted as
NIEM \cite{NIEM}, in that it uses non-equidistant mesh points, divides the
radial interval into partitions of adjustable size, and uses a very accurate
spectral integration technique involving Chebyshev polynomials. Because of its
inherent stability the S-IEM is likely the method of choice \cite{IEMA} for
situations requiring solutions out to large distances. The generalization of
this method so as to include the exchange potential \cite{baps}, \cite{KG} is
described further below.

In section 2 the basic equation to be solved will be described; in sections 3,
4 and 5 the S-IEM, the SVD-improved iterative method, and the method of
moments, respectively, will be reviewed; in section 6 the numerical comparison
between the four methods will be described; section 7 contains the summary and
conclusion; and Appendix 1 contains further details of the extension of the
S-IEM to the presence of exchange.

\section{Equations and Notations.}

The equation describing two electrons, one bound to a hydrogen-like nucleus of
charge Z, and another incident with kinetic energy $\overline{E}_{k}$\ on the
ground state of the atom is
\begin{equation}
-(\hbar^{2}/2\mu)\left[  \nabla_{\vec{r}_{1}}^{2}+\nabla_{\vec{r}_{2}}%
^{2}-\frac{Ze^{2}}{r_{1}}-\frac{Ze^{2}}{r_{2}}+\frac{e^{2}}{r_{12}}\right]
\Psi(\vec{r}_{1},\vec{r}_{2})=\overline{E}\;\Psi(\vec{r}_{1},\vec{r}_{2}),\;
\label{S}%
\end{equation}
where $\Psi(\vec{r}_{1},\vec{r}_{2})$ is the overall wave function,
$\overline{E}$ is the total energy, $e$ is the charge of the electron, $\mu$
is the is the reduced mass of the incident electron, $\hbar$ is Planck's
constant, $\vec{r}_{1}$ and $\vec{r}_{2}$ denote the position vectors of the
two electrons, $r_{1}$ and $r_{2}$ are the respective magnitudes, and
$r_{12}=|\vec{r}_{1}-\vec{r}_{2}|$ \ is the distance between the two
electrons. In order to transform the variables into atomic units, one
multiplies Eq.(\ref{S}) by $(2\mu/\hbar^{2})\,a_{0}^{2}$, where $a_{0}%
=(\hbar^{2}/\mu e^{2})$ is the Bohr unit of length, with the result
\begin{equation}
\left[  -\nabla_{\vec{x}_{1}}^{2}-\nabla_{\vec{x}_{2}}^{2}-\frac{2Z}{x_{1}%
}-\frac{2Z}{x_{2}}+\frac{2}{x_{12}}\right]  \Psi(\vec{r}_{1},\vec{r}%
_{2})=E\Psi(\vec{r}_{1},\vec{r}_{2}). \label{SA}%
\end{equation}
Here $\vec{x}=\vec{r}/a_{0}$ is a displacement vector in units of Bohr, and
$E=\overline{E}/\Re$ is the total energy in Rydberg units, with $\Re=\hbar
^{2}/(2\mu\,a_{0}^{2})$.

In the Hartree-Fock approximation one expands the total wave function in terms
of the bound states $\phi_{i},i=1,2,...$ of the atomic electron
\begin{equation}
\Psi(\vec{r}_{1},\vec{r}_{2})=\sum_{i}\left[  \psi_{i}(\vec{r}_{1})\;\phi
_{i}(\vec{r}_{2})\pm\psi_{i}(\vec{r}_{2})\;\phi_{i}(\vec{r}_{1})\right]  ,\;
\label{HF}%
\end{equation}
where $\psi_{i}$ are the wave functions of the scattered electron in channel
$i$ , to be determined from the solution of a set of coupled equations. The
$+$ or the $-$ signs occur for the spin singlet or triplet cases,
respectively. The subscript $i$ represents the set of all quantum numbers
which label the electron bound states. The corresponding principal quantum
number is $n_{i}$, and the corresponding bound state energy is $\varepsilon
_{i}=-(Z^{2}/n_{i}^{2})\Re$. The case of two or more bound electron states can
also be derived. The result is a set of coupled equations with local and
non-local pieces in the diagonal and off diagonal potentials. The latter are
semi-separable of fully separable, hence the method described here for the
one-channel case can also be applied. In the present study only the ground
state will be assumed, i.e., $i=1$ , and henceforth this subscript will be
dropped, and further, $Z=1$ . Under these assumptions the\ bound-state
electron energy is $\varepsilon=-\Re$ and the incident electron has the
asymptotic kinetic energy $\overline{E}_{k}=\overline{E}-\,\varepsilon$ .
Assuming that this is a positive quantity, the corresponding wave number $k$
in units of $a_{0}$ is given by
\begin{equation}
k^{2}=E_{k}=\left(  \overline{E}-\,\varepsilon_{i}\right)  /\Re=E-1. \label{E}%
\end{equation}

The equation for $\psi$ is obtained by truncating the sum in Eq. (\ref{HF}) to
one term, inserting it into Eq.\ (\ref{SA}), multiplying on the left by the
functions $\phi=\phi_{1}(\overrightarrow{r}_{2})$, and integrating over
$d^{3}r_{2}$. In the present numerical study only the case of orbital angular
momenta = 0 will be considered. The result is
\begin{equation}
(-\nabla_{x_{1}}^{2}+V(x_{1})-k^{2})\psi(\vec{r}_{1})\pm\left[  (-k^{2}%
+\varepsilon)\right]  \,<\phi|\psi>+<\phi|\frac{2}{x_{12}}\psi>\,=0, \label{1}%
\end{equation}
where $V=-\frac{2Z}{x_{1}}+<$ $\phi|\frac{2}{x_{12}}\phi>$ and the symbol
$<|>$ denotes $<A|B>\,=\,\int A(\vec{r}_{2})B(\vec{r}_{2})\,d^{3}r_{2}.$ If,
furthermore, the radial wave functions $R_{L}$ are introduced in the usual
way
\begin{equation}
\psi(\vec{r}_{1})=\frac{1}{r_{1}}\sum_{L}\;i^{L}(2L+1)R_{L}(r_{1})P_{L}%
(\cos\theta_{1}),
\end{equation}
where the the $P_{L}$ 's are Legendre Polynomials, and if Eq. (\ref{1}) is
multiplied by $P_{L}(\cos\theta_{1})$ and integrated over the solid angle
$d\Omega_{1}$ one obtains the final equation for the radial function $R_{0}$
for $L=0$
\begin{equation}
\left[  \frac{d^{2}}{dx_{1}^{2}}+k^{2}\right]  \,R_{0}(x_{1})=V(x_{1}%
)\,R_{0}(x_{1})\pm\int_{0}^{\infty}\mathcal{\Im}(x_{1},x_{2})\,R_{0}%
(x_{2})\,dx_{2}. \label{NL1}%
\end{equation}

In the above, (assuming $Z=1),$
\begin{align}
V(x)  &  =-2e^{-2x}(1+\frac{1}{x}),\label{V}\\
\Im(x_{1},x_{2})  &  =u(x_{1})\,u(x_{2})\left[  \gamma+\frac{2}{x_{12}%
}\right]  ,\label{I}\\
\gamma &  =-k^{2}-1,\;\label{G}\\
u(x)  &  =2xe^{-x},\;\label{u}\\
v(x)  &  =\frac{1}{x}u(x)=2e^{-x}. \label{v}%
\end{align}
The result for $u$ arises from the well known expression for $\phi_{1}$%
\[
\phi_{1}=\left(  Z/a_{0}\right)  ^{3/2}2\exp(-Zx_{2})\,Y_{00}(\vec{r}_{2})
\]
with $Z=1.$\ Utilizing the expansion of $1/x_{12}$ into Legendre Polynomials
in the angle between the directions of $x_{1}$ and $x_{2}$ , and remembering
that only the term in $P_{0}$ enters in the present case, one can recast the
kernel $\Im$ in the \emph{semi-separable} form
\begin{align}
\Im(x_{1},x_{2})  &  =2v(x_{1})u(x_{2})+\gamma\,u(x_{1})\,u(x_{2}%
)\;\;for\,x_{2}<x_{1}\;\label{fg}\\
\Im(x_{1},x_{2})  &  =2u(x_{1})v(x_{2})+\gamma\,u(x_{1})\,u(x_{2}%
)\;\;for\,x_{2}>x_{1}.\; \label{gf}%
\end{align}

The above equations (\ref{fg}) and (\ref{gf}) are the ones which will be used
by the three methods of calculation, to be described in sections 3, 4, and 5 below.

It is interesting to note that Eq. (\ref{NL1}) can be replaced by an
equivalent set of coupled equations\cite{CEQX}, \cite{WOT}%

\begin{align}
\left[  \frac{d^{2}}{dx^{2}}+k^{2}-V(x)\right]  \,R_{0}(x)  &  =\pm
V_{12}(x)\varphi_{2}(x)\pm c\;\gamma\,u(x)\nonumber\\
\left[  \frac{d^{2}}{dx^{2}}\right]  \varphi_{2}(x)  &  =V_{21}(x)R_{0}%
(x)\nonumber\\
c  &  =\int_{0}^{\infty}u(x^{\prime})R_{0}(x^{\prime})dx^{\prime}\;\;\;\;\;\;
\label{CE1}%
\end{align}
with
\begin{equation}
V_{12}(x)=V_{21}(x)=\sqrt{8}\exp(-x). \label{CE2}%
\end{equation}

The reason that it is possible here to replace a nonlocality by an equivalent
added channel is that the Green's function which corresponds to the operator
$d^{2}/dx^{2}$ is given by the product $f(x_{<})g(x_{>})$ , with $f(x)=x$ and
$g(x)=1.$ This equivalence is due to the fact that \cite{WOT} the Coulomb
interaction $1/r_{12},$ which appears in the first nonlocal term in $\Im,$ is
closely related to the Laplacian $d^{2}/dx^{2}.$ For angular momenta $L$ other
than zero\ it is sufficient to add the term $L(L+1)/x^{2}$ into the square
bracket of the second equation above. Whether the addition of extra channels
is feasible for exchange interactions different from the Coulomb interaction
remains to be investigated.

The above equations (\ref{CE1}) are a set of inhomogeneous coupled equations
which can be solved by conventional numerical means. \ However for more
general semi-separable nonlocalities, which cannot be reduced to a set of
equivalent coupled equations, the methods of solving Eq. (\ref{NL1}) presented
in the next sections can be used. The question of whether an exchange
nonlocality\ gives effects which are similar to a nonlocality due to coupling
to inelastic channels has been examined by many authors. For certain nuclear
scattering cases the two nonlocalities gave quite different
results\cite{lipperheide}. It is easy to understand the difference from Eqs.
(\ref{CE1}) above, since in the exchange case the second channel contains no
energy, and a inhomogeneous term is present, while both features are absent in
the inelastic coupled channel case.

\section{The Integral Equation Method (S-IEM)}

In this section we describe how the previously developed spectral integral
equation method for local potentials \cite{IEM} can be extended so as to
include the exchange terms. We drop the channel subscripts, since the equation
being discussed contains only one channel, and for ease of notation we replace
the function $R_{0}(x)$ in equation (\ref{NL1}) by $\varphi(x).$ In the
integral equation method S-IEM \cite{IEM}, the differential equation
(\ref{NL1}) is transformed into the equivalent integral equation
\begin{equation}
\varphi(x)=\sin(kx)+\int_{0}^{\infty}\mathcal{G(}x,x^{\prime})V(x^{\prime
})\varphi(x^{\prime})dx^{\prime}+\int_{0}^{\infty}\mathcal{F(}x,x^{\prime
\prime})\varphi(x^{\prime\prime})dx^{\prime\prime}. \label{ieq1}%
\end{equation}
where $\mathcal{G}$\ is the undistorted Green's function corresponding to the
momentum $k$,
\begin{align}
\mathcal{G(}x,x^{\prime})  &  =-\frac{1}{k}\cos(kx)\,\sin(kx^{\prime
})\;\;\;\;for\;x^{\prime}<x\nonumber\\
\mathcal{G(}x,x^{\prime})  &  =-\frac{1}{k}\sin(kx)\,\cos(kx^{\prime
})\;\;\;\;for\;x^{\prime}>x, \label{G0}%
\end{align}
and the kernel $\mathcal{F}$\ results from the presence of the exchange terms,
and is equal to the convolution of the Green's function with the nonlocal
potential $\Im,$ defined in Eqs. (\ref{fg}) and (\ref{gf}),
\begin{equation}
\mathcal{F(}x,x^{\prime\prime})=\int_{0}^{\infty}\mathcal{G(}x,x^{\prime}%
)\Im(x^{\prime},x^{\prime\prime})\,dx^{\prime}. \label{CV}%
\end{equation}

For a general nonlocal potential $\Im$ the kernel $\mathcal{F}$ is not
semi-separable. In this latter case the solution of the integral equation can
still be performed and gives rise to matrices which, although not sparse, have
a structure such that they can still be evaluated economically \cite{KKR}.
However, if the nonlocality $\Im$ is semi-separable, as is the case when it
results from exchange terms, then it can be shown \cite{baps}, \cite{KG}, that
the kernel $\mathcal{F}$ , Eq. (\ref{CV}) also is semi-separable and is of
rank 2. Even though the resulting expression for $\mathcal{F}$ is not as
simple as the rank 1 expression (\ref{G0}) for $\mathcal{G}$ , the
conventional IEM method for local potentials can be extended to this case, as
will be shown below. The advantage of this technique is that the ''big''
matrix, which occurs in the process of piecing together the local solutions
obtained for each partition, is a sparse band limited matrix, and hence the
complexity of the calculation remains proportional to the number of partitions
$m,$ rather than being of power $m^{3},$ as would be the case with general
non-sparse matrices. However, the complexity of the calculation also contains
a factor which increases like the cube of the number of bands, which, in the
case of the presence of nonlocal exchange potential doubles compared to the
local case, and hence the complexity of the calculation increases by an
overall factor of eight.

The semi-separable form of the kernel $\mathcal{F}$ is obtained by inserting
into Eq. (\ref{CV}) the expression (\ref{G0}) for the Green's function, and
using for $\Im(x^{\prime},x^{\prime\prime})$ the expressions given by Eqs.
(\ref{fg}) and (\ref{gf}). If one also combines the integral over
$\mathcal{G}V$ in Eq. (\ref{ieq1}) with the kernel $\mathcal{F}$ into a single
kernel $K$
\begin{equation}
\varphi(x)=\sin(kx)+\int_{0}^{\infty}K\mathcal{(}x,x^{\prime\prime}%
)\varphi(x^{\prime\prime})dx^{\prime\prime}, \label{ieq2}%
\end{equation}
one obtains for $K$ the result
\begin{align}
\emph{K}\mathcal{(}x,x^{\prime\prime})  &  =f_{1}(x)\,g_{1}(x^{\prime\prime
})+f_{2}(x)\,g_{2}(x^{\prime\prime})\;\;\;\;\;for\;x^{\prime\prime}<x\\
K\mathcal{(}x,x^{\prime\prime})  &  =p_{1}(x)\,q_{1}(x^{\prime\prime}%
)+p_{2}(x)\,q_{2}(x^{\prime\prime})\;\;\;for\;x^{\prime\prime}>x. \label{R2}%
\end{align}
The subscript 1 and 2 stands for the first and second semi-separable terms in
the rank-2 expression for $K$ , respectively. The functions $f$, $g$, $p$ and
$q$ are given by
\begin{align}
f_{1}(x)  &  =\cos
(kx)\;\;\;\;\;\;\;\;\;\;\;\;\;\;\;\;\;\;\;\;\;\;\;\;\;\;\;\;\;\;\;\;\;\;\;\;\;\;\;\;\;\;\;\;\;\;\;
\label{pe1}\\
g_{1}(x)  &  =2\mathcal{I}_{su}(x)\,v(x)-2\mathcal{I}_{sv}(x)\,u(x)-\frac
{1}{k}\sin(kx)\,V(x)\;\;\;\;\;\;\;\;\;\;\;\label{qe1}\\
f_{2}(x)  &  =\cos(kx)\left[  2\mathcal{I}_{sv}(x)+\gamma\mathcal{I}%
_{su}(x)\right]  \,+\sin(kx)\left[  2\mathcal{I}_{cv}(x)+\gamma\mathcal{I}%
_{cu}(x)\right] \\
g_{2}(x)  &
=u(x)\;\;\;\;\;\;\;\;\;\;\;\;\;\;\;\;\;\;\;\;\;\;\;\;\;\;\;\;\;\;\;\;\;\;\;\;\;\;\;\;\;\;\;\;\;\;\;\;\;\;
\end{align}
\begin{align}
p_{1}(x)  &  =\sin
(kx)\;\;\;\;\;\;\;\;\;\;\;\;\;\;\;\;\;\;\;\;\;\;\;\;\;\;\;\;\;\;\;\;\;\;\;
\label{pa1}\\
q_{1}(x)  &  =2\mathcal{I}_{cv}(x)\,u(x)-2\mathcal{I}_{cu}(x)\,v(x)-\frac
{1}{k}\cos(kx)\,V(x)\label{qa1}\\
p_{2}(x)  &  =\cos(kx)\mathcal{I}_{su}(x)+\sin(kx)\mathcal{I}_{cu}%
(x)\,\;\;\;\;\;\label{pa2}\\
q_{2}(x)  &  =2v(x)+\gamma
u(x),\;\;\;\;\;\;\;\;\;\;\;\;\;\;\;\;\;\;\;\;\;\;\;\;\;\;\;\;\;\;
\end{align}
where the functions $\mathcal{I}$ are defined by
\begin{align}
\mathcal{I}_{su}\mathcal{(}x)  &  =-\frac{1}{k}\int_{0}^{x}\sin
(kr)\,u(r)\,dr\nonumber\\
\mathcal{I}_{sv}\mathcal{(}x)  &  =-\frac{1}{k}\int_{0}^{x}\sin
(kr)\,v(r)\,dr\nonumber\\
\mathcal{I}_{cu}\mathcal{(}x)  &  =-\frac{1}{k}\int_{x}^{\infty}%
\cos(kr)\,u(r)\,dr\nonumber\\
\mathcal{I}_{cv}\mathcal{(}x)  &  =-\frac{1}{k}\int_{x}^{\infty}%
\cos(kr)\,v(r)\,dr \label{Iyz}%
\end{align}
and the functions $u$\ and $v$ are defined in Eqs. (\ref{u}) and (\ref{v}), respectively.

\subsection{The discretization.}

The integral equation to be solved is
\[
{\phi}(x)+\frac{f_{1}(x)}{k}\int_{0}^{x}g_{1}(x^{\prime}){\phi}(x^{\prime
})dx^{\prime}+\frac{f_{2}(x)}{k}\int_{0}^{x}g_{2}(x^{\prime}){\phi}(x^{\prime
})dx^{\prime}%
\]
\[
+\frac{p_{1}(x)}{k}\int_{x}^{\infty}q_{1}(x^{\prime}){\phi}(x^{\prime
})dx^{\prime}+\frac{p_{2}(x)}{k}\int_{x}^{\infty}q_{2}(x^{\prime}){\phi
}(x^{\prime})dx^{\prime}=sin(kx).
\]
It can be written in concise form as
\begin{equation}
(I+K)\,{\phi}(x)=sin(kx). \label{m1}%
\end{equation}
The radial distance $x$ is contained in the range $0\leq x\leq r_{\max}\,,$
\ where the upper limit $r_{\max}$ is chosen sufficiently large so that beyond
$r_{\max}$ the integrands can be neglected. We start by dividing the interval
$[0,r_{\max}]$ into $m$ \ partitions $[b_{0},b_{1}],\ [b_{1},b_{2}],$
$\dots,[b_{i-1},b_{i}],$ $\dots\lbrack b_{m-1},b_{m}],$ \ \ where the points
$b_{i}$ are not necessarily equispaced, similarly to what was done in
\cite{IEM}. Next we show that in each interval $i$ the global solution $\phi$
can be found as a linear combination of four local solutions of equation
(\ref{m1}) restricted to each of the subintervals of partition. Let $K_{i}$
denote the operator $K$ restricted to act only in the subinterval
$[b_{i-1},b_{i}].$ For example, $K_{i}$ \ operating on the function $\eta$
\ is given by
\begin{align*}
(K_{i}{\eta})(x)  &  =\frac{f_{1}(x)}{k}\int_{b_{i-1}}^{x}g_{1}(x^{\prime
}){\eta}(x^{\prime})dx^{\prime}+\frac{p_{1}(x)}{k}\int_{x}^{b_{i}}%
q_{1}(x^{\prime}){\eta}(x^{\prime})dx^{\prime}\\
&  +\frac{f_{2}(x)}{k}\int_{b_{i-1}}^{x}g_{2}(x^{\prime}){\eta}(x^{\prime
})dx^{\prime}+\frac{p_{2}(x)}{k}\int_{x}^{b_{i}}q_{2}(x^{\prime}){\eta
}(x^{\prime})dx^{\prime},\ \ \\
b_{i-1}  &  \leq x\leq b_{i}.
\end{align*}
Then, in terms of $K_{i}$, equation (\ref{m1}) can be rewritten as
\begin{align}
(I+K_{i}){\phi}(x)  &  =A^{(i)}p_{1}(x)+B^{(i)}p_{2}(x)+C^{(i)}f_{1}%
(x)+D^{(i)}f_{2}(x),\label{ne}\\
\ \ \ \ b_{i-1}  &  \leq x\leq b_{i},\nonumber
\end{align}
where use has been made of the fact that $p_{1}(x)=sin(kx).$ This result can
be obtained (see Ref. \cite{IEM}) by decomposing the integrals in equation
(\ref{m1}) into three domains: $[0,b_{i-1}]$, $[b_{i-1},b_{i}]$, and
$[b_{i},r_{\max}].$The second domain \ gives rise to the operator $K_{i}.$
Accordingly the constants\ are given by
\begin{equation}
A^{(i)}=1-\frac{1}{k}\int_{b_{i}}^{r_{\max}}q_{1}(x^{\prime}){\phi}(x^{\prime
})dx^{\prime}, \label{eq:q7}%
\end{equation}
\begin{equation}
B^{(i)}=-\frac{1}{k}\int_{b_{i}}^{r_{\max}}q_{2}(x^{\prime}){\phi}(x^{\prime
})dx^{\prime}, \label{eq:q8}%
\end{equation}
\begin{equation}
C^{(i)}=-\frac{1}{k}\int_{0}^{b_{i-1}}g_{1}(x^{\prime}){\phi}(x^{\prime
})dx^{\prime}, \label{eq:q9}%
\end{equation}
\begin{equation}
D^{(i)}=-\frac{1}{k}\int_{0}^{b_{i-1}}g_{2}(x^{\prime}){\phi}(x^{\prime
})dx^{\prime}, \label{eq:q10}%
\end{equation}
which are later found from the solution of matrix equation (\ref{Eq:B}) We
next define four functions $y_{i},z_{i},{\mu}_{i},{\xi}_{i}$ in each
subinterval $i$ by
\begin{align}
(I+K_{i})y_{i}(x)  &  =p_{1}(x),\label{eq:q5}\\
(I+K_{i})z_{i}(x)  &  =f_{1}(x),\\
(I+K_{i}){\mu}_{i}(x)  &  =p_{2}(x),\\
(I+K_{i}){\xi}_{i}(x)  &  =f_{2}(x).
\end{align}
In view of the fact that the operator $K_{i}$ is linear, the solution ${\phi
}(x)$ of equation (\ref{ne}) in each subinterval $i$ is given by
\begin{equation}
{\phi}(x)=A^{(i)}y_{i}(x)+B^{(i)}{\mu}_{i}(x)+C^{(i)}z_{i}(x)+D^{(i)}{\xi}%
_{i}(x),\;\;\hspace{3mm}b_{i-1}\leq x\leq b_{i}. \label{m2}%
\end{equation}
This result allows one to relate the constants $A,B,C,D$ in subinterval $i$
with those in other subintervals $j$, by inserting (\ref{m2}) into equations
(\ref{eq:q7})-(\ref{eq:q10}). The resulting equations, described in Appendix
1, can be transformed into a block tridiagonal-system, similarly to what was
done in our previous work, \cite{IEM},
\begin{equation}
\left[
\begin{array}
[c]{cccccc}%
\mathbf{I} & \mathbf{U}_{12} & \mathbf{0} & \mathbf{...} & ... & \mathbf{0}\\
\mathbf{U}_{21} & \mathbf{I} & \mathbf{U}_{23} & \mathbf{0} & \mathbf{...} &
\mathbf{...}\\
\mathbf{0} & \mathbf{U}_{32} & \mathbf{I} & \mathbf{U}_{34} & \mathbf{0} &
\mathbf{...}\\
& \ddots & \ddots & \ddots & \ddots & \\
\mathbf{0} & \mathbf{...} &  &  & \mathbf{I} & \mathbf{U}_{m-1,m}\\
&  &  &  & \mathbf{U}_{m,m-1} & \mathbf{I}%
\end{array}
\right]  \left[
\begin{array}
[c]{c}%
{\bar{\mathbf{\Delta}}}_{1}\\
{\bar{\mathbf{\Delta}}}_{2}\\
{\bar{\mathbf{\Delta}}}_{3}\\
\vdots\\
\\
{\bar{\mathbf{\Delta}}}_{m}%
\end{array}
\right]  =\left[
\begin{array}
[c]{c}%
{\bar{\mathbf{\emph{0}}}}\\
{\bar{\mathbf{\emph{0}}}}\\
\vdots\\
\vdots\\
{\bar{\mathbf{\emph{0}}}}\\
{\bar{\mathbf{\emph{E}}}}%
\end{array}
\right]  \label{Eq:B}%
\end{equation}
\medskip\ where the quantities ${\bar{\mathbf{\Delta}}},$ ${\bar
{\mathbf{\emph{0}}}}$, and ${\bar{\mathbf{\emph{E}}}}$ are 1$\times4$ column
vectors\medskip\
\begin{align*}
{\bar{\mathbf{\Delta}}}_{i}  &  =[A^{(i)},B^{(i)},C^{(i)},D^{(i)}{]}^{T},\\
{\bar{\mathbf{\emph{0}}}}  &  =[0,0,0,0]^{T},\\
{\bar{\mathbf{\emph{E}}}}  &  =[1,0,0,0]^{T}%
\end{align*}
and each block is a $4\times4$ matrix,
\begin{equation}
\mathbf{U}_{i,i+1}=\left[
\begin{array}
[c]{cccc}%
(\alpha_{1}y)_{i+1}-1 & (\alpha_{1}\mu)_{i+1} & (\alpha_{1}z)_{i+1} &
(\alpha_{1}\xi)_{i+1}\\
(\alpha_{2}y)_{i+1} & (\alpha_{2}\mu)_{i+1}-1 & (\alpha_{2}z)_{i+1} &
(\alpha_{2}\xi)_{i+1}\\
0 & 0 & 0 & 0\\
0 & 0 & 0 & 0
\end{array}
\right]
\end{equation}
\begin{equation}
\mathbf{U}_{i+1,i}=\left[
\begin{array}
[c]{cccc}%
0 & 0 & 0 & 0\\
0 & 0 & 0 & 0\\
(\beta_{1}y)_{i} & (\beta_{1}\mu)_{i} & (\beta_{1}z)_{i}-1 & (\beta_{1}%
\xi)_{i}\\
(\beta_{2}y)_{i} & (\beta_{2}\mu)_{i} & (\beta_{2}z)_{i} & (\beta_{2}\xi
)_{i}-1
\end{array}
\right]  \label{Eq:14}%
\end{equation}
\medskip for $i=1,...,m-1,$ and $\mathbf{I}$ is a $4\times4$ identity matrix.
The entries of the matrices are integrals in each partition $i$ of products of
the known functions $q_{1},q_{2},g_{1},g_{2},$ \ and the numerically computed
functions $y,\mu,z,$ \ and $\xi$ , where by definition,\
\[
\left(  pq\right)  _{i}=\int_{b_{i-1}}^{b_{i}}p(x)\,q(x)\,dx.
\]
It is noteworthy that the structure of the matrix in Eq.(\ref{Eq:B}) is very
similar to the structure encountered for a set of four coupled channels (see
Eq. (34) in Ref. \cite{IEM}. (This reference contains further details of the
discretization technique). The system of equations (\ref{Eq:B}) is solved by
Gaussian elimination specialized for band limited matrices, see e.g.
\cite{GOLUB}. Its complexity is $4mp(p+1)-\frac{2}{3}p^{3},$ where $p$\ is the
number of non-zero subdiagonals, (band-width), and $m$ is the number of
partitions. In Eq. (\ref{Eq:B}) $p=7$. Since the number of grid points per
partition, $16,$ is larger than $p=7,$ it is clear that the overall cost will
be dominated by the cost of solving Eqs. (\ref{eq:q5}) in all partitions,
which is of order $16^{3}m$\ 

\section{The SVD-improved Iterative Method.}

The first of our three comparison methods consists in replacing the exchange
kernel by a small number of fully separable terms, and carrying out iterations
only over the remainder, as will be described in this section, and as is given
with more detail in Ref. \cite{essaid}. As is well known, the Green's function
for a Schr\"{o}dinger equation with both local and non local but fully
separable potentials can be obtained without much difficulty by adding terms
to the Green's function distorted only by the local potential. By contrast,
the older iterative method of taking the nonlocal kernel into account
perturbatively consists in writing Eq. (\ref{NL1}) in the form
\begin{equation}
\left[  \frac{d^{2}}{dx_{1}^{2}}-V(x_{1})+k^{2}\right]  \,R_{0}(x_{1})=\pm
\int_{0}^{\infty}\mathcal{\Im}(x_{1},x_{2})\,R_{0}(x_{2})\,dx_{2} \label{NL2}%
\end{equation}
and then transforming it into the iterative integral equation
\begin{equation}
R_{0}^{(n+1)}(x_{1})=f(x_{1})+\int_{0}^{\infty}\mathcal{G}_{V}(x_{1}%
,x^{\prime})\;\left(  \pm\int_{0}^{\infty}\mathcal{\Im}(x^{\prime}%
,x_{2})\,R_{0}^{(n)}(x_{2})\,dx_{2}\right)  dx^{\prime}. \label{IT1}%
\end{equation}
In the above, $f(x)$ is the ''regular'' solution of
\begin{equation}
\left[  \frac{d^{2}}{dx_{1}^{2}}-V(x_{1})+k^{2}\right]  \,f(x)=0, \label{HO}%
\end{equation}
and $\mathcal{G}_{V}(x_{1},x^{\prime})$ is the Green's function which
corresponds to the left hand side of Eq. (\ref{NL2}). It is distorted by the
local potential $V,$ and can be expressed in terms of semi-separable
expressions involving two independent solutions $f(x)$ and $g(x)$ of Eq.
(\ref{HO}),
\begin{align}
\mathcal{G}_{V}\mathcal{(}x,x^{\prime})  &  =-\frac{1}{k}\,f(x)g(x^{\prime
})\;\;\;for\,\,\;\;x\leq x^{\prime}\label{eq:green}\\
\mathcal{G}_{V}\mathcal{(}x,x^{\prime})  &  =-\frac{1}{k}\,g(x)f(x^{\prime
})\;\;\;for\,\,\;\;x>x^{\prime},
\end{align}
as is well known. The functions $f$ and $g$ are normalized such that their
Wronskian is equal to $k$. The iteration is started by using the solution in
the absence of the exchange terms for the first $(n=0)$ guess $R_{0}%
^{(0)}=f(x_{2}).$

The rate of convergence of the iterations depends on the norm of
\[
\mathcal{F}_{V}\mathcal{(}x,x^{\prime\prime})=\int_{0}^{\infty}\mathcal{G}%
_{V}\mathcal{(}x,x^{\prime})\Im(x^{\prime},x^{\prime\prime})\,dx^{\prime}.
\]
This norm in turn depends on the norm of $\Im$, and on the norm of
$\mathcal{G}_{V}.$ The latter becomes large at small incident energies $k^{2},
$ in view of the presence of the factor $1/k$ in Eq. (\ref{eq:green}), and
hence the iteration will diverge for a sufficiently small value of $k. $ The
rate of convergence also depends on the $\pm$ sign in front of the exchange
integrals, as was found in the numerical examples described below. This effect
does not occur in the other methods described in this paper because the latter
do not make use of the iteration on $\mp\mathcal{F}_{V}$.

In what follows in this section, we describe a method, to be denoted as SVD,
which reduces the norm of the nonlocal kernel $\Im$ by decomposing it into a
sum of a fully separable kernel of low rank plus a remainder. The separable
terms are placed in the left hand side of Eq. (\ref{NL2}), the Green's
function in the presence of both the local distorting potential $V$ and the
separable nonlocal pieces of the kernel is obtained, and hence iterations of
the form of Eq. (\ref{IT1}) can be carried out, where $\Im$ is now the
residual kernel. This way the ''$k$''-divergence can be shifted to smaller
values of $k,$ but it cannot be avoided.

\subsection{The separable content of the nonlocal kernel.}

The singular value decomposition method (SVD) \cite{SVD} is used to decompose
the kernel $\mathcal{\Im}(x_{1},x_{2})$ into a number of fully separable terms
plus a remainder. The method is as follows. First a numerical integration
algorithm is chosen which divides the range of integration $[0,R_{\max}]$ into
a set of $N$ discrete points. Correspondingly the kernel $\mathcal{\Im}%
(x_{1},x_{2})$ is transformed into a $N\times N$ matrix $K(i,j)$, with
$i,j=1,2,..N.$ We next perform a singular value decomposition on $K$. The SVD
method is based on a theorem of linear algebra according to which any $M$
$\times$ $N$ matrix $K$ can be written as the product of an $M$ $\times$ $M$
orthogonal matrix $U$, an $M$ $\times$ $N$ diagonal matrix $\Sigma$ with
positive or zero elements, and the transpose of an $N$ $\times$ $N$ orthogonal
matrix $V$. ( A matrix $U$ is orthogonal if $UU^{T}=U^{T}U=I,$ which means
that its columns are normalized and orthogonal to each other, and so are the
rows.) For our purpose it is sufficient to consider the case $N=M$. In this
case we can rigorously write
\begin{equation}
K=U\,\Sigma\,V^{T}=\sum_{s=1}^{N}\,\sigma_{s}\mathbf{u}_{s}\mathbf{v}_{s}^{T}
\label{e20}%
\end{equation}
where the columns of $U$ and $V$ are the column vectors $\mathbf{u}_{s},$ and
$\mathbf{v}_{s}$, respectively, and $\Sigma$ is a diagonal matrix of the
non-negative quantities $\sigma_{s},$ $s=1,2,...N,$ ordered by decreasing size
(the largest ones first). The latter are the ''singular values''. As a result
of the above, a fully separable piece of rank $n$ can be separated out of the
matrix $K,$ leaving a residual matrix $K^{R},$%
\begin{equation}
K=K^{S}+K^{R}.
\end{equation}
by carrying the sum in Eq.(\ref{e20}) to a upper limit $n$ which includes only
the largest values $\sigma_{s}.$
\begin{equation}
K^{S}(i,j)=\sum_{s=1}^{n}u_{js}\,\sigma_{s}\text{ }v_{si}\;,\,\;or\;\;\;K^{S}%
=\sum_{s=1}^{n}\,\sigma_{s}\mathbf{u}_{s}\mathbf{v}_{s}^{T}\equiv\sum
_{s=1}^{n}\mathbf{u}_{s}\rangle\,\sigma_{s}\text{ }\langle\mathbf{v}_{s}.
\label{KS1}%
\end{equation}
The last entry into the above equation uses the Dirac notation for a vector
and its transpose. The remainder $K^{R}$ is given by
\begin{equation}
K^{R}=K-K^{S}=\sum_{s=n+1}^{N}\sigma_{s}\,\mathbf{u}_{s}\text{ }\mathbf{v}%
_{s}^{T}. \label{KR1}%
\end{equation}
\medskip

\subsection{Greens function for a separable potential.}

In order to obtain the Green's function $\mathcal{G}_{V+K^{s}}\mathcal{(}%
x,x^{\prime}),$ which is distorted by both the local potential V and the fully
separable Kernel $K^{S},$ we rewrite Eq. (\ref{IT1}) symbolically in the form
\begin{equation}
\psi(x)=f(x)+\mathcal{G}_{V}\mathcal{(}K^{S}+K^{R})\psi\label{G1}%
\end{equation}
where the integration over the variables is implicitly assumed. For
simplicity, let us assume that only two terms in $K^{S}$ are responsible for
the divergence of the iterative Green's function approach, Eq. (\ref{IT1}). In
order to obtain the overlap integrals $\langle v_{i}\psi\rangle,i=1,2$ we
multiply Eq. (\ref{G1}) on the left with $\sqrt{\sigma_{i}}\langle v_{i}$ and
integrate over all $x$'s, with the result that $\sqrt{\sigma_{i}}\langle
v_{i}\psi\rangle=$ $\sqrt{\sigma_{i}}\langle v_{i}f\rangle+\sqrt{\sigma_{i}%
}\langle v_{i}\mathcal{G}_{V}(K^{S}+K^{R}\rangle.$ Rearranging terms one
obtains the following matrix equation for $\sqrt{\sigma_{i}}\langle v_{i}%
\psi\rangle$%
\begin{equation}
M\sqrt{\sigma}\left[
\begin{array}
[c]{c}%
\langle v_{1}\psi\rangle\\
\langle v_{2}\psi\rangle
\end{array}
\right]  =\sqrt{\sigma}\left[
\begin{array}
[c]{c}%
\langle v_{1}f\rangle\\
\langle v_{2}f\rangle
\end{array}
\right]  -\sqrt{\sigma}\left[
\begin{array}
[c]{c}%
\langle v_{1}K^{R}\psi\rangle\\
\langle v_{2}K^{R}\psi\rangle
\end{array}
\right]  , \label{G2}%
\end{equation}
where
\[
M=\left(
\begin{array}
[c]{cc}%
1+\mathcal{G}_{11} & \mathcal{G}_{12}\\
\mathcal{G}_{21} & 1+\mathcal{G}_{22}%
\end{array}
\right)  ,\;\;\;\;\sqrt{\sigma}=\left(
\begin{array}
[c]{cc}%
\sqrt{\sigma_{1}} & 0\\
0 & \sqrt{\sigma_{2}}%
\end{array}
\right)
\]
and
\[
\mathcal{G}_{ij}=\sqrt{\sigma_{i}}\langle v_{i}\mathcal{G}_{V}\,u_{j}%
\rangle\sqrt{\sigma_{j}},\;\;\;\;\;i=1,2.
\]
Solving Eq. (\ref{G2}) for $\left[  \langle v_{1}\psi\rangle,\langle v_{2}%
\psi\rangle\right]  $ and inserting the result into Eq. (\ref{G1}), one
obtains
\begin{equation}
\psi=f-\mathcal{G}_{V}\,\left[  u_{1}\rangle,u_{2}\rangle\right]  \sqrt
{\sigma}M^{-1}\sqrt{\sigma}\left\{  \left[  \langle v_{1}f\rangle,\langle
v_{2}f\rangle\right]  ^{T}-\left[  \langle v_{1}K^{R}\psi\rangle,\langle
v_{2}K^{R}\psi\rangle\right]  ^{T}\right\}  , \label{G3}%
\end{equation}
from which the result for $\mathcal{G}_{V+K^{s}}$ emerges:
\begin{equation}
\mathcal{G}_{V+K^{s}}=\mathcal{G}_{V}\left\{  1-\left[
\begin{array}
[c]{cc}%
u_{1}\rangle & u_{2}\rangle
\end{array}
\right]  \sqrt{\sigma}M^{-1}\sqrt{\sigma}\left[
\begin{array}
[c]{c}%
\langle v_{1}\\
\langle v_{2}%
\end{array}
\right]  \right\}  \label{G4}%
\end{equation}

The numerical result for the triplet phase shift, shown in Table 1 of section
6, used five sets of singular value functions $\mathbf{u}_{s}$ and
$\mathbf{v}_{s}$ and required five iterations of Eq. \ref{IT1}. Without the
use of the SVD expansion, the iterations did not converge for $k\leq0.3.$ For
values of $k<0.1$ the iterations using the SVD expansion did not converge for
either the singlet or triplet cases.

\section{The Modified Integral Equation Method (M-IEM)}

In this section we describe the method proposed by Kim and
Udagawa\cite{udagawa}, which we call the modified integral equation method
(M-IEM). The method is well documented in the literature, and hence only a
brief description is given here. It starts from the following equation
obtained by rewriting Eq.(\ref{NL1});
\begin{equation}
\left[  \frac{d^{2}}{dx_{1}^{2}}-V(x)+k^{2}\right]  \,\varphi(x)=\pm
\lambda(x)\text{ } \label{NL3}%
\end{equation}
\begin{equation}
\lambda(x)=\int_{0}^{\infty}\mathcal{\Im}(x,x^{\prime})\,\varphi(x^{\prime
})\,dx^{\prime}.\; \label{AX}%
\end{equation}
We then transform the equation into the integral form as
\begin{equation}
\varphi(x)=\varphi^{(0)}(x)\pm\int_{0}^{\infty}\mathcal{G}^{\prime
}(x,x^{\prime\prime})\lambda(x^{\prime\prime})dx^{\prime\prime},\;
\label{ieq3}%
\end{equation}
where $\varphi^{(0)}(x)$ and $\mathcal{G}^{\prime}(x,x^{\prime})$ satisfy
\begin{align}
\left[  \frac{d^{2}}{dx_{1}^{2}}-V(x)+k^{2}\right]  \,\varphi^{(0)}(x)  &
=0\text{ }\label{deq1}\\
\left[  \frac{d^{2}}{dx_{1}^{2}}-V(x)+k^{2}\right]  \,\mathcal{G}^{\prime
}(x,x^{\prime\prime})  &  =\delta(x-x^{\prime\prime})\text{ } \label{deq2}%
\end{align}
Further, we modify Eq.(\ref{ieq3}) by multiplying both sides by $\mathcal{\Im
}(x,x^{\prime})$ and carrying out the integration over $x^{\prime}$. The
result is
\begin{align}
\lambda(x)  &  =\lambda^{(0)}(x)\,\pm\,\int_{0}^{\infty}\int_{0}^{\infty
}\mathcal{\Im}(x,x^{\prime})\mathcal{G}^{\prime}(x^{\prime},x^{\prime\prime
})\lambda(x^{\prime\prime})\;dx^{\prime\prime}dx^{\prime},\label{ieq4}\\
\lambda^{(0)}(x)  &  =\int_{0}^{\infty}\mathcal{\Im}(x,x^{\prime}%
)\varphi^{(0)}(x^{\prime})dx^{\prime}\; \label{lam0}%
\end{align}
The equation we solve is (\ref{ieq4}). Since both $\varphi^{(0)}(x)$ and
$\mathcal{G}^{\prime}(x,x^{\prime\prime})$ are defined in terms of the local
potential $V(x)$, they can be calculated without any problem. This means that
once the solution $\lambda(x)$ of Eq.(\ref{ieq4}) is obtained, then
$\varphi(x)$ can be calculated from Eq. (\ref{ieq3}).

In solving Eq.(\ref{ieq4}), use is made of the Lanczos method \cite{white}. It
is worth noting that the application of the Lanczos method for solving
Eq.(\ref{ieq4}) is possible, since $\lambda(x)$ is a bounded function, as can
be seen from the fact that it is essentially given in terms of the bounded
nonlocal potential function $\mathcal{\Im}(x,x^{\prime})$. This makes it
possible to expand $\lambda(x)$ in terms of an orthonormal set of functions,
as is done in Eq.(\ref{expand}) below. This is not the case for $\varphi(x)$
in Eq.(\ref{ieq3}), since $\varphi(x)$ is not bounded.

We first expand $\lambda(x)$ in terms of the orthonormal set of functions
$D_{i}(x)$ with $i=0,1,2,.....,N_{i}$ which are generated as follows:
\begin{align}
D_{0}(x)  &  =\frac{1}{d_{0}}\lambda^{(0)}(x),\;\;\;\;\label{base0}\\
D_{i}(x)  &  =\frac{1}{d_{i}}\left(
\begin{array}
[c]{c}%
\int_{0}^{\infty}\int_{0}^{\infty}\mathcal{\Im}(x,x^{\prime})\mathcal{G}%
^{\prime}(x^{\prime},x^{\prime\prime})D_{i-1}(x^{\prime\prime})\,dx^{\prime
\prime}dx^{\prime}-\\
\sum_{j=0}^{i-1}D_{j}(x)\alpha_{j\;i-1}%
\end{array}
\right)  , \label{base2}%
\end{align}
with
\begin{equation}
a_{ji}=\left\{
\begin{array}
[c]{c}%
\int_{0}^{\infty}\int_{0}^{\infty}\int_{0}^{\infty}\tilde{D}_{j}%
(x)\mathcal{\Im}(x,x^{\prime})\mathcal{G}^{\prime}(x^{\prime},x^{\prime\prime
})D_{i}(x^{\prime\prime})\,dx^{\prime\prime}dx^{\prime}dx,\;\;j\leq i+1\\
\\
0\;\;\;\;\;\;\;\;\;\;\;\;\;\;\;\;\;\;\;\;\;\;\;\;\;\;\;\;\;\;\;\;\;\;\;\;\;\;\;\;\;\;\;\;\;\;\;\;\;\;\;\;j>i+1
\end{array}
\right\}  \label{alpha}%
\end{equation}
The normalization constant $d_{i}$ in Eqs.(\ref{base0}) and (\ref{base2}) is
determined from the condition
\begin{equation}
\int_{0}^{\infty}\tilde{D}_{i}(x)D_{i}(x)dx=1 \label{norm}%
\end{equation}
$\tilde{D}_{i}(x)$ being the conjugate function to $D_{i}(x)$. The
coefficients $\alpha_{ji}$ given by Eq.(\ref{alpha}) are those determined from
the usual Schmidt orthonormalization procedure. Now we write $\lambda(x)$ as
\begin{equation}
\lambda(x)=\sum_{j=0}^{N_{i}}C_{j}D_{j}(x), \label{expand}%
\end{equation}
where $C_{j}$ are the expansion coefficients.

Inserting Eq.(\ref{expand}) into Eq.(\ref{ieq4}), one can easily derives a set
of inhomogeneous linear equations for the expansion coefficients $C_{j}$,
i.e.,
\begin{equation}
\sum_{j}(\delta_{ij}-\alpha_{ij})C_{j}=d_{0}\delta_{0i}. \label{coeff}%
\end{equation}
The values of $C_{j}$ are then determined by solving Eq.(\ref{coeff}). Note
that Eq.(\ref{coeff}) can be solved rather easily, because $\alpha_{j\,i}=0$
for $j>i+1$ (see Eq.(\ref{alpha})). In addition, the value of $N_{i}$ can be
chosen as a small number. This helps greatly in making the actual numerical
calculations very fast.

\bigskip

\section{Numerical Results}

%

\begin{figure}
[ptb]
\begin{center}
\includegraphics[
height=3.5699in,
width=5.6014in
]%
{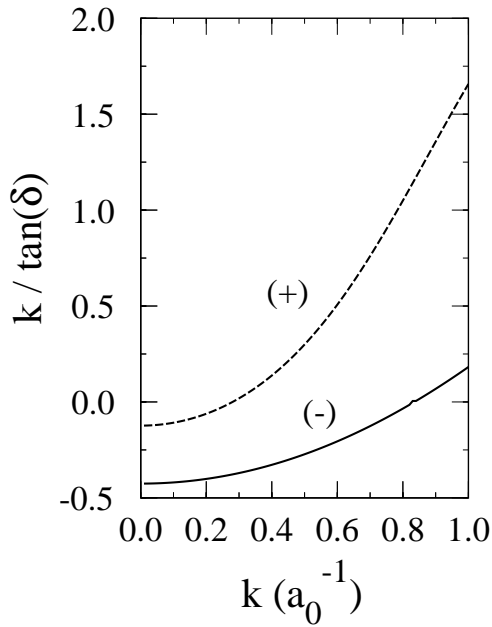}%
\caption{Dependence of the singlet (+) and triplet (-) $L=0$ phase shifts on
the incident wave number, displayed in the form $k/\tan(\delta)$. Where these
curves pass through zero, $\tan(\delta)$ goes through infinity, as is shown in
the next two figures. Neither the M-IEM or the S-IEM had any difficulty
evaluating these quantities either for the small values of $k$ or in the
vicinity of the zeros.}%
\label{FIG1}%
\end{center}
\end{figure}
The bench-test calculation performed by the three methods described above
consists in obtaining the $L=0$ phase shift for the scattering of an electron
from the ground-state of an Hydrogen atom, in the presence of exchange terms,
both for the singlet and the triplet states, $\delta^{(+)}$ and $\delta^{(-)}$
, respectively. The methods are the S-IEM, the SVD, and the M-IEM. The older
integral equation method is denoted as NIEM (the ''N'' stands for
non-iterative), and a representative result is taken from the paper by Sams
and Kouri \cite{kouri}, since these authors describe their accuracy for
exactly the same test case as ours. The SVD, M-IEM and the NIEM methods use
equi-spaced mesh points, since their auxiliary functions are the solutions of
local differential equations using finite difference methods, while the S-IEM,
as mentioned above, uses non-equispaced mesh points, which are the zeros of a
Chebyshev polynomial of a certain order (16 in this case), in each of the
partitions into which the radial interval is decomposed.%
\begin{figure}
[ptbptb]
\begin{center}
\includegraphics[
height=3.5699in,
width=5.6014in
]%
{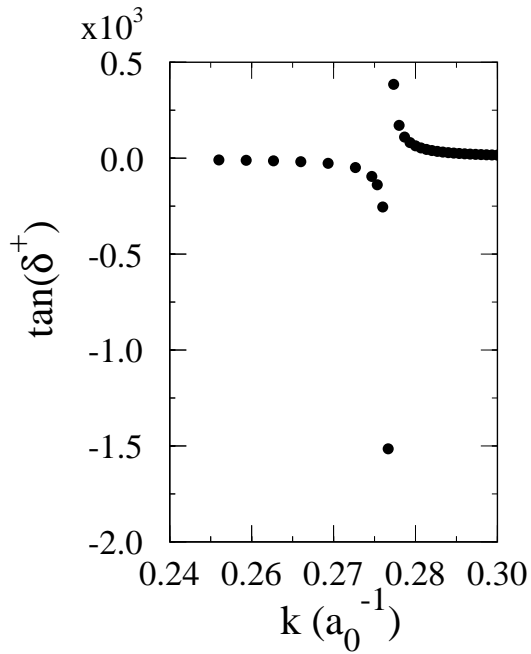}%
\caption{Momentum dependence of the singlet phase shift in the vicinity of
$\pi/2$ (mod. $\pi$), as calculated by the S-IEM. Please note the scale of the
y-axis.}%
\label{FIG2}%
\end{center}
\end{figure}

\begin{figure}
[ptb]
\begin{center}
\includegraphics[
height=3.5699in,
width=5.6014in
]%
{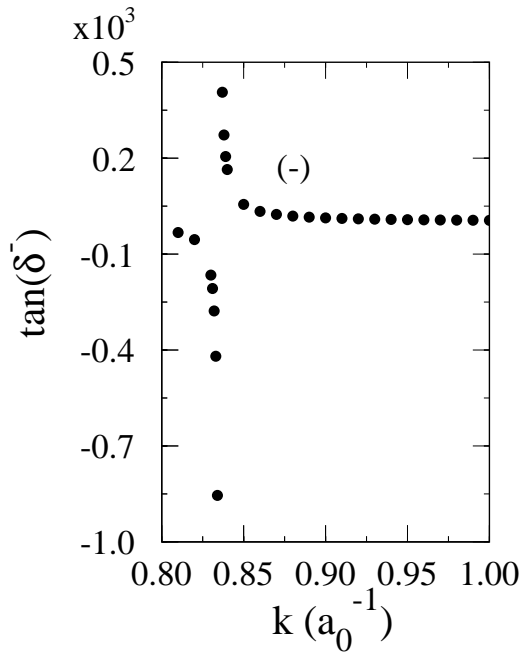}%
\caption{Same as Fig. 2 for the triplet phase shift.}%
\label{FIG3}%
\end{center}
\end{figure}

The $k$ dependence of the phase shifts is shown in Fig. 1, by plotting the
ratio $k/\tan(\delta).$ When either of the two curves crosses the $0$ line,
the corresponding value of $\tan(\delta)$ becomes infinite,as is shown in
Figs. (\ref{FIG2}) and (\ref{FIG3}), and the respective phases shift have the
value $\pi/2,$ modulus $\pi.$ Both the M-IEM and the S-IEM methods had no
difficulty in reproducing the singularity in $\tan(\delta),$ and both were
able to reach arbitrarily small values of the momentum $k.$ By contrast, the
SVD method could not obtain results for $k<1.0(a_{0})^{-1}.$%

Since well documented accuracy studies exist for the NIEM \cite{kouri} we
examined the rate of convergence of the phase shift as a function of number
the mesh points for a case which is treated in Ref. \cite{kouri}. The case
chosen is the singlet phase shift with exchange, $\mathbf{\delta}^{(+)}$. The
value of the wave number is $k=0.2\,(a_{0})^{-1}$ , and the maximum radial
distance is $20\,a_{0}.$ The number of significant figures obtained for each
of the non-spectral methods, all using a mesh size of 0.005 $a_{0},$ is shown
in the three last rows of Table I. The result for the spectral method S-IEM,
also shown. \medskip

\begin{center}
Table 1: Accuracy of $\mathbf{\delta}^{(+)}$ for various algorithms.\medskip
\\[0pt]\medskip%
\begin{tabular}
[c]{|l|c|l|}\hline\hline
\textbf{Method} & $\mathbf{\delta}^{(+)}$ & \textbf{\# of Pts.}\\\hline\hline
S-IEM & \multicolumn{1}{|l|}{1.8701579} & \multicolumn{1}{|r|}{80}\\\hline
M-IEM$^{a)}$ & \multicolumn{1}{|l|}{1.870156} & \multicolumn{1}{|r|}{4000}%
\\\hline
NIEM$\,^{b)}$ & \multicolumn{1}{|l|}{1.87015} & \multicolumn{1}{|r|}{4000}%
\\\hline
SVD & \multicolumn{1}{|l|}{1.8701} & \multicolumn{1}{|r|}{4000}\\\hline
\end{tabular}
$\smallskip$

$^{a)}$ Five basis states $D$ are used in this calculation\\[0pt]$^{b)}%
$Non-iterative method of Ref. \cite{kouri}\medskip
\end{center}

\bigskip

The convergence of the four methods with the number of mesh-points is
illustrated in Fig. \ref{FIG4}. The number of significant figures for a given
number of mesh points is determined from the stability of the result obtained
after rounding, when compared to the result with the next higher number of
points. It is clear from the figure that the S-IEM method reaches higher
accuracy with a smaller number of points than the other methods shown. With
160 mesh points (the corresponding number of partitions is 10) the value
obtained for $\delta^{(+)}=1.87015788462442$ rad. is the same , to within the
quoted number of 15 significant figures, as the result for 224 mesh points.
This is close to machine accuracy, and shows that the accumulation of
round-off errors is small in the S-IEM method, confirming previous studies.
(See Fig. 1 in the 1997 paper quoted in Ref. \cite{IEM}). The discrepancy in
the seventh significant figure between the S-IEM and the M-IEM could be due to
the fact that only five basis functions were used \ for the latter. This point
has not been investigated further.

The scattering length $a$ and effective range $r_{e}$ for this one state
electron-hydrogen scattering calculation have also been examined. The
procedure is similar to the one used for a previous atom-atom scattering
bench-mark calculation \cite{IEMA}. It is based on the low momentum expansion
of the scattering phase shift
\begin{equation}
k\cot\delta_{0}=-\frac{1}{a}+r_{e}k^{2}+O(k^{3}). \label{scattl}%
\end{equation}
The left hand side of the above expression is calculated for two very small
values of the wave number $k$ , differing by a factor of two, and the values
of $a$ and $r_{e}$ are solved for. The procedure is repeated for decreasing
values of $k$ and increasing values of the maximum radial distance $r_{\max} $
and of the number of mesh-points until stability in the results is found to a
given number of significant figures. For the values of $a$ and $r_{e}$ listed
in the tables below for the\ S-IEM method, values of $k\approx10^{-5} $ ,
$r_{\max}\approx50$ and approximately $1000$ mesh points were found to be
adequate. However, contrary to what was done in Ref. \cite{IEMA}, the value of
$r_{\max}$ was not extrapolated to $\infty$ via a perturbative method. For the
M-IEM case, the values of $a$ and $r_{e}$ in Table IV are extracted from Eq.
\ref{scattl} by calculating $k\cot\delta_{0}$ for the two values of
$k=0.00001$ and $0.01,$ and for $T=20.$ Excellent agreement with the S-IEM
values is obtained.\medskip\medskip

\begin{center}
Table II: Scattering lengths $a$.\\[0pt]\medskip%

\begin{tabular}
[c]{|l||c|c|c|}\hline\hline
\textbf{Method} & \textbf{Singlet} & \textbf{No exchange} & \textbf{Triplet}%
\\\hline\hline
S-IEM & \multicolumn{1}{||l|}{8.100312397} &
\multicolumn{1}{|r|}{-9.44716668854} & \multicolumn{1}{|l|}{2.349396156}%
\\\hline
M-IEM & \multicolumn{1}{||l|}{8.1003} & \multicolumn{1}{|l|}{-9.44716} &
\multicolumn{1}{|l|}{2.3494}\\\hline
\end{tabular}

\end{center}

\medskip

\begin{center}
Table III: Effective Range $r_{e}$.\\[0pt]\medskip%

\begin{tabular}
[c]{|l||l|l|l|}\hline\hline
\textbf{Method} & \textbf{Singlet} & \textbf{No exch.} & \textbf{Triplet}%
\\\hline\hline
S-IEM & \multicolumn{1}{||l|}{1.51201} & \multicolumn{1}{|l|}{0.766797} &
0.6105\\\hline
M-IEM & \multicolumn{1}{||l|}{1.51} & 0.767 & 0.612\\\hline
\end{tabular}

\bigskip%
\begin{figure}
[ptb]
\begin{center}
\includegraphics[
height=3.6071in,
width=3.9548in
]%
{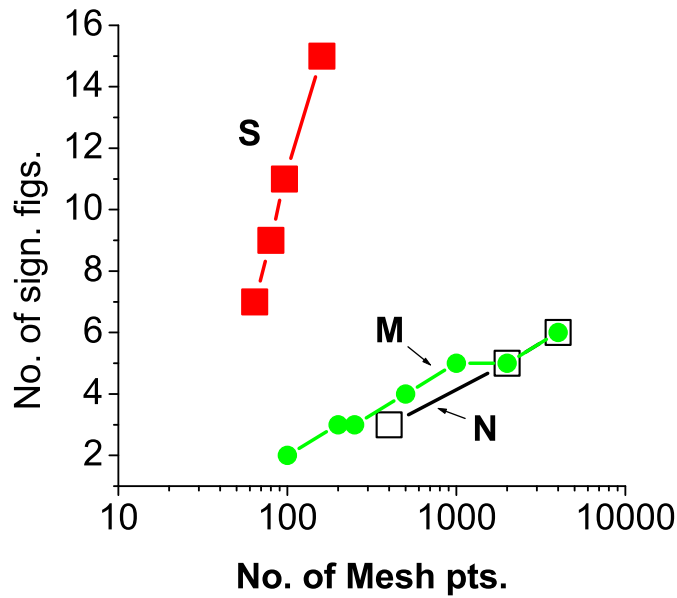}%
\caption{Comparison of the accuracy of the S-IEM, M-IEM and NIEM integral
equation methods for the calculation of the singlet phase shift, as a function
of the number of meshpoint used. The incident momentum is $k\,=0.2(a_{0}%
)^{-1},$ and the value of the radial cut-off point is $r_{\max}=20\,a_{0}.$
The NIEM results are taken from Ref. \cite{kouri}. The accuracy for a given
number of mesh points of each method is determined by the number of
significant figures which are stable (after rounding), as compared with the
result for the next higher number of meshpoints.}%
\label{FIG4}%
\end{center}
\end{figure}

\end{center}

\section{Summary and Conclusions.}

In this paper four methods were compared to solve the one-dimensional
Schr\"{o}dinger equation in the presence of the exchange nonlocality for the
case of electron scattering from a hydrogen atom, with only the lowest energy
state of the bound electron being included. The oldest method in the
literature proceeds by first solving the equation in the presence of only the
local potential, and then including the nonlocal part through Green's function
iteration. The iterations converge only for a limited range of parameters, and
one of the methods described here improves upon the convergence by separating
out of the nonlocal kernel a fully separable part by means of the Singular
Value Decomposition method (SVD). By this means the region of convergence
could be extended to a larger domain, but for our example convergence still
fails at small values of the momentum, $k<0.1(a_{0})^{-1}$ and the maximum
accuracy achieved was five significant figures. Another method was developed
in the literature, in which the differential Schr\"{o}dinger equation is first
transformed into an Lippman-Schwinger integral equation, and is then solved
non-iteratively (NIEM). In our example, taken from the literature
\cite{kouri}, this method achieved six significant figures of accuracy, but
appears not to work for small values of the incident momentum. Improved
accuracy and the viability for all values of $k$ was achieved in the present
study by extending a previously developed spectral solution of a
Lippman-Schwinger integral equation with local potentials \cite{IEM},
\cite{IEMA}, to the case with an exchange-type nonlocality. This extension was
possible because the exchange nonlocality is of a semi-separable character.
The resulting method (S-IEM) gives substantially higher accuracy (15
significant figures) than the NIEM, and converges much faster with the number
of mesh-point in the integration interval than the NIEM, as is illustrated in
Fig. 4. A fourth method, (M-IEM) developed previously for nonlocalities
occurring in nuclear physics \cite{udagawa} achieves seven figures of
accuracy, and has no difficulty in coping with small values of $k.$ The rate
of convergence of this method was comparable to that of the NIEM. The reason
is due to the fact that the auxiliary functions needed for both methods, as
well as the integration algorithms, are based on a finite difference
algorithm, whose error usually decreases inversely with the number of mesh
points according to a well defined power. From inspection of Fig. 4, this
power has the relatively low value of 2.5. Both the M-IEM and the SVD methods
have the advantage that they can be used for non-localities which are more
general than the semi-separable\ exchange ones. The S-IEM also can be applied
to these cases, but, at its present stage of development, the large matrix in
Eq. \ref{Eq:B} is then no longer sparse \cite{KKR}.

In summary, four methods of calculating the scattering phase shift in electron
atom collision were compared for a numerical test case, and the advantages and
disadvantages of each were discussed.

\newpage\medskip

{\Large Appendix 1}

The relation between all coefficients in (\ref{m2}) can be written in the
matrix form,
\begin{equation}
\left[
\begin{array}
[c]{cccc}%
M_{11} & M_{12} & M_{13} & M_{14}\\
M_{21} & M_{22} & M_{23} & M_{24}\\
M_{31} & M_{32} & M_{33} & M_{34}\\
M_{41} & M_{42} & M_{43} & M_{44}%
\end{array}
\right]  \left[
\begin{array}
[c]{c}%
\overline{\mathbf{A}}\\
\overline{\mathbf{B}}\\
\overline{\mathbf{C}}\\
\overline{\mathbf{D}}%
\end{array}
\right]  =\left[
\begin{array}
[c]{c}%
\overline{\mathbf{1}}\\
\overline{\mathbf{0}}\\
\overline{\mathbf{0}}\\
\overline{\mathbf{0}}%
\end{array}
\right]  \;\;\;eq:15 \label{eq:15}%
\end{equation}
where
\begin{align*}
{\bar{\mathbf{A}}}  &  =[A_{1},...,A_{m}]^{T}\hspace{5mm}{\bar{\mathbf{B}}%
}=[B_{1},...,B_{m}]^{T},\\
{\bar{\mathbf{C}}}  &  =[C_{1},...,C_{m}]^{T}\hspace{5mm}{\bar{\mathbf{D}}%
}=[D_{1},...,D_{m}]^{T},\\
{\bar{\mathbf{1}}}  &  =[1,...,1]^{T}\hspace{5mm}\overline{\mathbf{0}%
}=[0,...,0]^{T},
\end{align*}
and each of the block matrices $M_{ij}$ are either upper triangular matrices
$U$ or lower triangular matrices $L$ with either $1^{\prime}s$ or $0^{\prime
}s$ in the main diagonal, respectively (the subscripts are accordingly $1$ or
$0$ ). These matrices $M_{i,j}$ are thus of the form
\[
U_{1}=\left[
\begin{array}
[c]{ccccc}%
1 & \gamma_{2} & \gamma_{2} & \cdots & \gamma_{m}\\
0 & 1 & \gamma_{3} & \cdots & \gamma_{m}\\
\vdots &  & \ddots &  & \vdots\\
0 & \cdots & 0 & 1 & \gamma_{m}\\
0 & \cdots &  & 0 & 1
\end{array}
\right]  ,\;\;\;\;U_{0}=\left[
\begin{array}
[c]{ccccc}%
0 & \gamma_{2} & \gamma_{2} & \cdots & \gamma_{m}\\
0 & 0 & \gamma_{3} & \cdots & \gamma_{m}\\
\vdots &  & \ddots &  & \vdots\\
0 & \cdots & 0 & 0 & \gamma_{m}\\
0 & \cdots &  & 0 & 0
\end{array}
\right]
\]
or
\[
L_{1}=\left[
\begin{array}
[c]{ccccc}%
1 & 0 & \cdots &  & 0\\
\delta_{1} & 1 & \cdots &  & 0\\
\delta_{1} & \delta_{2} & 1 & \cdots & 0\\
\vdots & \vdots & \ddots & 1 & 0\\
\delta_{1} & \delta_{2} & \cdots & \delta_{m-1} & 1
\end{array}
\right]  ,\;\;\;\;L_{0}=\left[
\begin{array}
[c]{ccccc}%
0 & 0 & \cdots &  & 0\\
\delta_{1} & 0 & \cdots &  & 0\\
\delta_{1} & \delta_{2} & 0 & \cdots & 0\\
\vdots & \vdots & \ddots & 0 & 0\\
\delta_{1} & \delta_{2} & \cdots & \delta_{m-1} & 0
\end{array}
\right]  ,
\]
in which the entries $\gamma$ or $\delta$ are given in the table below\medskip

\begin{center}
{Table 1: Entries }$\gamma${\ and }$\delta${\ in the block matrices }$M{=}U
${\ or }$M=L.$%

\begin{tabular}
[c]{|c|c|}\hline
$M_{11}=U_{1}$ \ with ${\gamma}_{i}=(q_{1}y)_{i}$ & $M_{12}=U_{0}$ \ with
${\gamma}_{i}=(q_{1}{\mu})_{i}$\\\hline
$M_{13}=U_{0}$ \ with ${\gamma}_{i}=(q_{1}z)_{i}$ & $M_{14}=U_{0}$ \ with
${\gamma}_{i}=(q_{1}{\xi})_{i}$\\\hline
$M_{21}=U_{0}$ \ with ${\gamma}_{i}=(q_{2}y)_{i}$ & $M_{22}=U_{1}$ \ with
${\gamma}_{i}=(q_{2}{\mu})_{i}$\\\hline
$M_{23}=U_{0}$ \ with ${\gamma}_{i}=(q_{2}z)_{i}$ & $M_{24}=U_{0}$ \ with
${\gamma}_{i}=(q_{2}{\xi})_{i}$\\\hline
$M_{31}=L_{0}$ \ with ${\delta}_{i}=(g_{1}y)_{i}$ & $M_{32}=L_{0}$ \ with
${\delta}_{i}=(g_{1}{\mu})_{i}$\\\hline
$M_{33}=L_{1}$ \ with ${\delta}_{i}=(g_{1}z)_{i}$ & $M_{34}=L_{0}$ \ with
${\delta}_{i}=(g_{1}{\xi})_{i}$\\\hline
$M_{41}=L_{0}$ \ with ${\delta}_{i}=(g_{2}y)_{i}$ & $M_{42}=L_{0}$ \ with
${\delta}_{i}=(g_{2}{\mu})_{i}$\\\hline
$M_{43}=L_{0}$ \ with ${\delta}_{i}=(g_{2}z)_{i}$ & $M_{44}=L_{1}$ \ with
${\delta}_{i}=(g_{2}{\xi})_{i}$\\\hline
\end{tabular}

\end{center}

\noindent Using elementary row operations on equation (\ref{eq:15}) and then
changing the order of the variables, the coefficient matrix of equation
(\ref{eq:15}) can be transformed into the block tridiagonal system
(\ref{Eq:B}) to (\ref{Eq:14}), given in the text. Further details can be found
in Ref. \cite{IEM}. \clearpage
\end{document}